\documentclass[review,authoryear,3p]{elsarticle}
\usepackage{epsfig,graphics,amssymb,amsmath,subeqnarray,color}

\begin{document}

\title{Crawling scallop: Friction-based locomotion with one degree of freedom}
\author{Gregory L. Wagner\corref{cor1}}
\ead{glwagner@ucsd.edu}
\author{Eric Lauga\corref{cor2}}
\ead{elauga@ucsd.edu}
\ead[url]{http://maeresearch.ucsd.edu/lauga/}
\address{Department of Mechanical and Aerospace Engineering, University of California San Diego, 9500 Gilman Drive, La Jolla, CA 92093-0411, USA.}
\cortext[cor1]{Corresponding author.  Telephone:  (781) 710-0871.  Fax: (858) 822-3107.}
\cortext[cor2]{Telephone: (858) 822-7925}

\begin{abstract}
Fluid-based locomotion at low Reynolds number is subject to the constraints of the scallop theorem, which dictate that body kinematics identical under a time-reversal symmetry (in particular, those with a single degree of freedom) cannot display locomotion on average. The implications  of the theorem naturally compel one to ask whether similar symmetry constraints exist for locomotion in different environments.  In this work we consider locomotion along a surface where forces are described by isotropic Coulomb friction. To address whether motions with a single degree of freedom can lead to transport, we analyze a model system consisting of two bodies whose separation  distance undergoes periodic time variations.  The behavior of the two-body system is entirely determined by the kinematic specification of their separation, the friction forces, and  the mass of each body. We show that the constraints of the scallop theorem can be escaped in frictional media if two asymmetry conditions are met at the same time: the frictional forces of each body against the surface must be distinct and the time-variation of the body-body separation must vary asymmetrically in time (so quick-slow or slow-quick in the extension-contraction phases). Our results are demonstrated numerically and interpreted using asymptotic expansions.
\end{abstract}

\maketitle

%%%%%%%%%%%%%%%%%%
\section{Introduction}

The capacity for locomotion is essential for the survival of much of life on Earth and is manifested in strategies as diverse as the organisms which depend on it.  In fluids such as water or air and on scales ranging from microns to tens of meters, creatures swim and fly by beating flagella, tails, wings, undulating their bodies, or actuating pumps \citep{vogel1994}.  On land, animals crawl, walk, run, hop, climb, and slither using friction between their bodies and the ground, or undulate and appear to ``swim'' through sand or soil \citep{alexander2003}.

The physics and scale of an environment determine the scope of successful propulsive strategies.  The strategy employed by a scallop, for example, which is to quickly open its shell, displacing a large amount of fluid, and then close it slowly, displacing a small amount of fluid, is ineffective if attempted in a fluid environment where inertial forces are overwhelmed by viscous forces.  This notion forms the basis for the ``scallop theorem'', which holds that locomotion at low Reynolds number is not possible if the kinematics of the body are identical under a time-reversal symmetry -- which is always true if the deformation is controlled by a single degree of freedom  \citep{purcell1977}. The theorem relies on the linearity and time-independence of the equations of motion for the fluid (the Stokes equation) and states that in order to achieve self-propulsion, a body at low Reynolds number must deform in a manner indicating a clear direction of time, for example in a waving motion \citep{lauga09}. 

The beauty and important implications of this theorem, which greatly restrict the viable propulsive strategies available to organisms and machines at small Reynolds numbers, naturally compel one to seek similar fundamental results for other types of locomotion.  For example, consider terrestrial locomotion for which organisms make use of friction forces between their bodies and a surface.  Are strategies in this physical environment similarly constrained?

In general, strategies for terrestrial locomotion may be divided into two categories: those relying on the movement of limbs (walking, hopping, climbing, and running) and those relying on movements of the body (crawling and slithering).  A number of studies have explored slithering locomotion, including its mechanics \citep{home1812, mosauer1932, gray1946, hu2009}, evolutionary advantages \citep{gans1975}, energetic efficiency \citep{walton1991}, optimization \citep{jingAlben2012} as well as its application to robotic propulsion \citep{hirose1990, transeth2009}.  On crawling-types of locomotion, which is perhaps the simplest of all forms of terrestrial locomotion and is the subject of this work, investigations have examined the locomotion of maggots \citep{berriganPepin1995} and earthworms \citep{quillin1999}, as well as hypothetical discrete-mass systems including systems consisting of two masses \citep{chernousko2002, zimmerman2004} or a chain of three or more \citep{figurina2004,zimmerman2004,zimmerman2007b,zimmerman2009,bolotnik2011} connected by springs or rigid mechanisms and actuated by external forces or kinematic constraints.

A feature common to studies of the discrete-mass crawling strategies is that they consider systems with either many degrees of freedom \citep{zimmerman2004,zimmerman2007b,zimmerman2009,bolotnik2011}, anisotropic friction coefficients \citep{zimmerman2004}, or analyze only locomotion in the case where the masses alternately stick and slide \citep{chernousko2002}.  Here we consider the most basic situation where deformation with one degree of freedom (and therefore time-reversible in its sequence of shapes) actuates a system with isotropic Coulomb friction coefficients and which always interacts with the surface through sliding friction forces, and propose to quantify the minimal requirements necessary to achieve locomotion.  Is locomotion even possible with isotropic friction using time-reversible deformations?
What are the associated minimum necessary mechanical or kinematic properties? How do they determine the direction and magnitude of motion?

To answer these questions, we consider a simple system consisting of two bodies which rest on a flat surface and are joined by a mechanism that enforces the time-variation of their  separation distance.  The variation of in the distance between the bodies gives rise to friction forces which determine their motion. Friction forces are assumed to be isotropic (independent of the direction of body velocity) as it is evident that friction anisotropy will trivially lead to locomotion. We demonstrate that locomotion is possible in a system with only one degree of freedom provided that there is an asymmetry both in the friction properties of the system  as well as the deformation kinematics (meaning the time-periodic variations in length display a quick-slow or slow-quick sequence of extension-contraction strokes).  Finally, we develop a qualitative physical explanation for the mechanics of strategy of locomotion through an asymptotic analysis of the equations of motion \citep{benderOrszag1999} and present a brief exploration of the space of physical parameters available to the system.

%%%%%%%%%%%%%%%%%%%%%%%%%%%%
\section{Mathematical description of the two-body system}

\begin{figure}[t!]
\centering
\includegraphics[width = 0.7\textwidth]{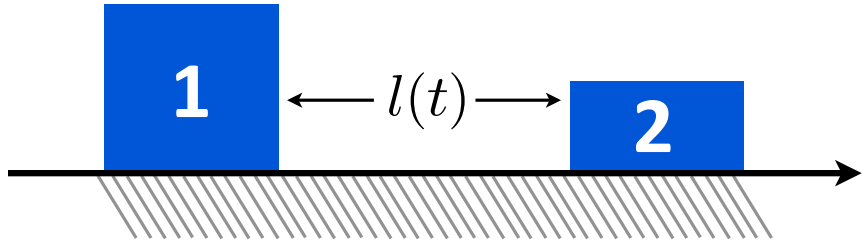}
\caption{Two-body system (1 and 2) with prescribed relative distance, $l(t)$, translating along the $x$ direction. }
\label{setup}
\end{figure}

Consider a system consisting of two bodies with position $x_i$ ($i=1,2$), velocity $\dot{x}_i$, acceleration  $\ddot{x}_i$, and mass $m_i$ which are supported by a flat surface and are constrained to translate along a horizontal line (the $x$ direction).  The bodies are connected by a mechanism which prescribes their relative distance and is the single degree of freedom available to the system to form the basis for its locomotion.  The prescribed relative position of the bodies is denoted $l(t) = x_2 - x_1$.  A schematic representation of this system is shown in Fig.~\ref{setup}.

The variation  in time of $l(t)$ gives rise to a driving force, $F_{G,i}$, which is exerted on each body by the linkage and is opposed by a friction force, $F_{f,i}$, exerted on each body by the solid surface.  The motion of the bodies is described by Newton's second law, 
\begin{equation}
\ddot{x}_i = \frac{1}{m_i} \left ( F_{G,i} + F_{f,i} \right ).
\label{eq:eom}
\end{equation}
The friction force may be described by a variety of different models.  Here, we restrict our analysis to isotropic Coulomb friction forces which are independent of the sliding velocity.  We use a standard model which introduces a ``static'' friction force that requires that the force opposing it be greater in order for the given body to translate, which we take to be the same as the sliding friction force.  Mathematically we may express this by writing the friction law
\begin{equation}
F_{f,i} = \left\{ 
  \begin{array}{l l}
    - F_i \text{ sgn}(\dot{x}_i) & \quad \text{if $| F_{G,i} | > F_i$,}\\
    - F_{G,i} & \quad \text{if $| F_{G,i} | \le F_i$,}\\
  \end{array} \right.
\label{frictionForce}
\end{equation}
where $F_i$ is then the magnitude of the sliding friction force exerted on body $i$ by the surface.  Here we choose to represent the physical properties of the bodies and the surface as a friction force as opposed to a friction coefficient in order to simplify the analysis; friction coefficients $\mu_i$ can be found by computing $\mu_i = F_i / m_i g$.  Without loss of generality we assume that $F_1/m_1 > F_2/m_2$ (or $\mu_1 > \mu_2$) to ensure that only body 1 sticks to the surface if the driving force is too weak to overcome the friction of both  bodies.

For the prescription of the relative distance between the  bodies,  we assume that $l(t)$ is a periodic function which consists of an extension phase in which the bodies are pushed apart and a contraction phase in which the bodies are pulled together.  Again without loss of generality, we specify the kinematics such that the extension phase is followed by the contraction phase and the bodies begin and end each period of the kinematic specification at their minimum separation.  If the extension phase and contraction phase are opposite and equal in magnitude, we say that the kinematic specification is ``symmetric''; otherwise the kinematic specification is ``asymmetric''.  

In order to determine the driving force between the bodies arising from the variation of their relative distance,  we subtract the acceleration of each  body to obtain
\begin{equation}
\ddot{x}_1 - \ddot{x}_2 = \ddot l(t) = \frac{1}{m_2} \left ( F_{f,2} + F_{G,2} \right ) - \frac{1}{m_1} \left ( F_{f,1} + F_{G,1} \right ).
\end{equation}
The relative position is enforced by a rigid mechanism which implies that $F_{G,1} = - F_{G,2}$, and so we have
\begin{equation}
F_{G_2} = \frac{m_1 m_2}{m_1 + m_2} \left ( \ddot{l} - \frac{F_{f, 2}}{m_2} + \frac{F_{f, 1}}{m_1} \right ).
\label{drivingForce}
\end{equation}
Inserting this relation into the equations of motion for each  body,  we find
\begin{equation}
\begin{split}
\ddot{x}_1 &= - \frac{m_2}{m_1 + m_2} \ddot{l} + \frac{F_{f_2} + F_{f_1}}{m_1 + m_2}, \\
\ddot{x}_2 &= \frac{m_1}{m_1 + m_2} \ddot{l} + \frac{F_{f_2} + F_{f_1}}{m_1 + m_2},
\label{blockMotion}
\end{split}
\end{equation}
which completes our mathematical description of the body motion.  For this equation of motion to hold for the blocks, the driving force must be large enough (and the kinematic acceleration large enough) such that both blocks slide.  When the kinematic acceleration is too small for both blocks to slide, only the block with the smaller friction force will move.  Also, since the kinematics are specified externally, the driving force is necessarily always large enough to force motion in one of the blocks.  We therefore find that the Froude number, ${\rm Fr}$,  which is the ratio between driving forces and friction forces and is defined as ${\rm Fr} = (m_1+m_2) L / T^2 (F_1+F_2)$, must be relatively large.  In contrast, the locomotion of snakes is usually associated with ${\rm Fr} < 1$ \citep{hu2009}, and while insufficient data is available to calculate $\rm Fr$ for organisms which use closely related mechanisms (such as maggots, see \cite{berriganPepin1995}), they are almost certainly associated with low Froude numbers as well.  It seems likely that the reason for this is tied to the energy costs associated with maintaining high Froude numbers in frictional media.

The most striking feature of the equations of motion is the discontinuous dependence of the friction force on the body velocity.  This discontinuity is the critical nonlinearity which enables locomotion.  However, while the equations are globally nonlinear, it is apparent that they are linear on intervals for which the direction of body motion does not change and that within these the position and velocity of the bodies can be found analytically.  The full solution can therefore be determined by pasting together the exact solutions for adjacent intervals.  Another technique to find steady solution computes the position and velocity of the system for a single period and equates initial and final conditions.

%%%%%%%%%%%%%%%
\section{Locomotion with one degree of freedom is possible}

\begin{figure}[t!]
\centering
\includegraphics[width = 0.9\textwidth]{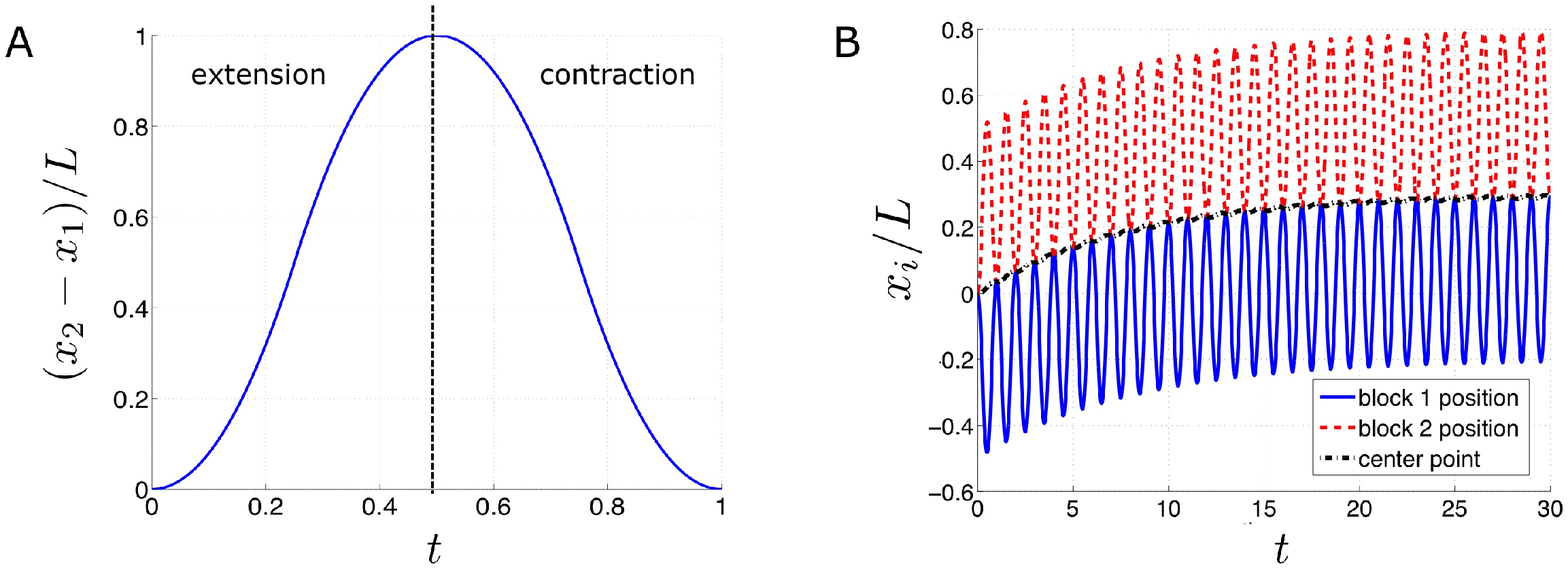}
\includegraphics[width = 0.9\textwidth]{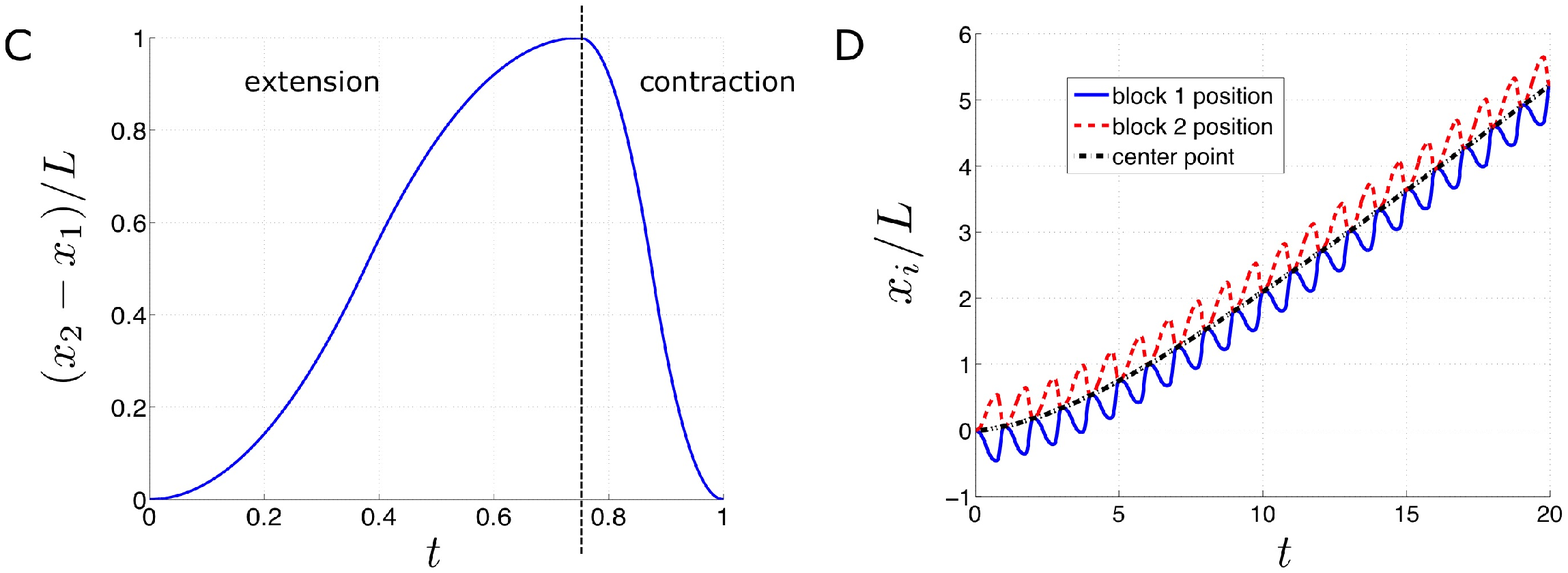}
\caption{Example of asymptotic stationary oscillation for equal masses  ($m_1 = m_2 = 1$) but unequal friction forces 
 ($F_1 = 4$, $F_2 = 1$) and  a symmetric kinematic specification.  
(A) Positions of both bodies (body 1, thick blue line; body 2, dashed red line) and geometric center (dash-dotted black line).  
(B) Time-variation of the normalized separation distance between the two bodies, $(x_2-x_1)/L$,  showing symmetric kinematics.
(C) and (D): same as in (A) and (B) but with asymmetric kinematics leading to locomotion.}
\label{translationExample_osc}
\end{figure}

With the mathematical description of body motion established, we ask whether sliding locomotion with a single degree of freedom is even possible.  For our simple two-body system we observe two distinct modes of asymptotic motion, as illustrated in Fig.~\ref{translationExample_osc}:  stationary oscillation around a fixed point (no locomotion, see Fig.~\ref{translationExample_osc} A and B) and net translation (locomotion, see Fig.~\ref{translationExample_osc} C and D).  In both examples from  Fig.~\ref{translationExample_osc}, bodies 1 and 2 have the same mass ($m_1=m_2$)  but are subject to different friction forces on the surface ($F_2\neq F_1$).  The two examples  only differ in their kinematic specification: in A and B, the bodies are actuated with a symmetric kinematic specification, resulting in stationary oscillation, whereas in C and D an asymmetric kinematic specification leads to locomotion.  In both cases the bodies exhibit transient behavior at the onset of motion which progresses into an asymptotic, stable mode of motion after long times.

In general, three different regimes exist in the choice of relative friction force and kinematic specification:  (a) equal friction forces and arbitrary kinematics, (b) unequal friction forces and symmetric kinematics, and (c) unequal friction forces and asymmetric kinematics.  Regimes (a) and (b) lead to stationary oscillation, whereas parameter choices falling into regime (c) lead to translation. Locomotion with a single degree of freedom is therefore possible provided that there is an asymmetry in both the kinematics (actuation occurs at a different rate one way than the other) and the friction forces exerted by each block on the surface.

\section{Physics of  locomotion with one degree of freedom: double symmetry-breaking}

The locomotion of the system depends on the frictional symmetry between the bodies as well as the symmetry between the extension and contraction phases of the kinematic specification.  Physically, the asymmetry in friction forces is required because if the friction forces are symmetric, a solution exists in which the velocities of the bodies always oppose one another, and accordingly their friction forces cancel each other such that their resulting equations of motion are merely proportional to the kinematic specification.  The asymmetry in the kinematic specification is required because if the extension and contraction phases are opposite and equal in magnitude, a solution exists for which the net motion of the bodies over each phase is equal and opposite, which cumulatively result in no net motion over the kinematic period.  The breaking of each of these symmetries divides the space of possible system properties into the three aforementioned fundamental regimes; we now examine each regime in detail.

\subsection{Symmetric friction forces}\label{A}

When the friction forces exerted by each bodies are identical, the system does not translate.  If the bodies are pushed apart in opposing directions after an initial time $t = t_0$, the velocity and position of body 1 are given by
\begin{eqnarray}
\displaystyle\dot{x}_1 &=& - \frac{m_2}{m_1 + m_2} \dot{l},\label{6} \\
\displaystyle x_1 &= &- \frac{m_2}{m_1 + m_2} \left [ l(t) - l(t_0) \right ] + x_1(t_0),
\end{eqnarray}
where we have used $\dot{x}_1(0) = \dot{l}(0) = 0$.  The position of body 2 can be found similarly, and if for simplicity we use $x_C(0) = 0$, we find the position of the center of mass to be
\begin{equation}
x_C(t) = \frac{1}{2} \left ( \frac{m_1}{m_2} - 1 \right ) \left [ x_1(t) - x_1(t_0) \right ].
\end{equation}
Since $x_1(t)$ is periodic in time (via Eq.~\ref{6}), we find that the position of the geometric center is periodic as well and the system exhibits stationary oscillation.  When $m_1 = m_2$,  the center of the  system is stationary for all times.

\subsection{Asymmetric friction forces, symmetric kinematics}\label{B}

Further, even if the friction forces exerted by the bodies are asymmetric, the system will not translate if the kinematic specification is symmetric.  This is best described mathematically by stating that $l(t) = l(T - t)$ for all $t$, where $T$ is the period of oscillation of $l(t)$. In this case, we are able to find a solution in which the system does not exhibit locomotion.  To simplify the argument we consider the extension and contraction phases separately, which we denote $A$ and $B$ respectively.  Each phase begins at a time $t_0$ and ends at time $t_f$, and the position of the geometric center during each phase is $x_A$ and $x_B$.

When the friction forces exerted by the bodies are unequal and the system is oriented as in Fig.~\ref{setup}, we observe that the system velocity increases over the course of an extension phase.  It is then possible to find a small negative initial system velocity at the beginning of the extension phase such that $-\dot{x}_A(t_0) = \dot{x}_A(t_f)$.  Then, by examination of the equations of motion (\ref{eq:eom}), it is evident that if the relative acceleration imposed by the contraction phase is opposite that imposed by the extension phase, then $\dot{x}_B(t_0) = \dot{x}_A(t_f)$ implies $\dot{x}_B(t_0) = -\dot{x}_B(t_f)$ and $\dot{x}_A(t_0) = \dot{x}_B(t_f)$, and the considered initial system velocity constitutes an equilibrium initial system velocity in a periodic solution for the system motion.  Further, the same logic applies to the system position: if the kinematic specification is opposite in sign and equal in magnitude, and if the body velocities are opposite in sign, then the motion of the geometric center over the extension and contraction phases, respectively, will be opposite in sign and equal, and the system will experience no net motion cumulatively over both phases.  Locomotion is therefore not possible when the extension and contraction phases of the kinematic specification are symmetric, even under asymmetric friction. The statements of \S \ref{A} and \S \ref{B} constitute thus a form of scallop theorem for locomotion under isotropic friction.

\subsection{Asymmetric friction forces, asymmetric kinematics}

We have demonstrated that asymmetric friction forces and asymmetric kinematics can result in locomotion, but the question remains: why?  Some insight can be gained by considering the system in the asymptotic limit of zero kinematic period, or fast kinematics and strong driving forces.  We make two key observations.  First, over the duration of an extension phase, the body with a smaller friction force (body 2) will move farther forward ($x>0$) than the body with a smaller friction force (body 1) when the bodies are pushed part.  This implies trivially that the geometric center of the system will move forward during an extension phase and backwards during a contraction phase. Second, if the extension phase is longer (and the kinematic velocity faster) than the contraction phase, the net change in system position over both intervals will be positive.  This can be understood by examining the first terms of an asymptotic expansion in the extension phase duration for the change in system velocity and position over the phase. Due to the piecewise nature of the equations, we must consider separately the case where the bodies begin the extension phase at rest from the case where they begin with some initial velocity.

\subsubsection{A single extension phase when bodies are initially at rest}

\begin{figure}[t]
\centering
\includegraphics[width = .95\textwidth]{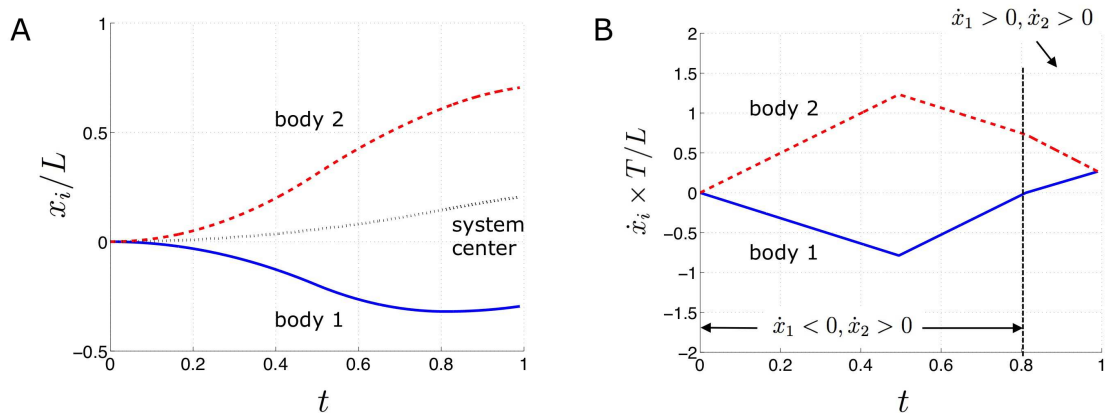}
\caption{Illustration of body position (A) and velocity (B) in an extension phase when both bodies start from rest; (A): normalized body positions, $x_i/L$;  (B): normalized body velocities, $\dot{x}_i \times (T/L)$.  There are two intervals distinguished by the signs of the body velocities; in the first, the bodies have opposite velocity and are pushed apart; in the second, both bodies translate forward.}
\label{onePhaseIntervals}
\end{figure}

Consider the two-block system when it is initially at rest at time $t_0 = 0$.  In this case, the extension phase is divided into two intervals distinguished by the sign of the velocity of body 1, as  illustrated in Fig.~\ref{onePhaseIntervals}  where we show both the body positions (A) and velocities (B). In the first interval, which constitutes the greater part of the extension phase, the bodies are pushed apart from their stationary position so that body 2 moves forward and body 1 backwards, and $\ddot{l}(t) >  0$.  This interval ends when the velocity of body 1, which is lesser in magnitude than body 2 due to its larger friction force, reaches  zero and changes direction.  In the second interval, both bodies move forward until the end of the interval is reached. 

The motion of each body is described by Eq.~\eqref{blockMotion}.  In the first interval we have that $F_{f_1} = F_1$ and $F_{f_2} = -F_2$, which yields
\begin{equation}
\ddot{x}_1 = - \frac{m_2}{m_1+m_2} \ddot{l}(t) + \frac{F_1 - F_2}{m_1 + m_2}\cdot
\end{equation}
Integrating from $t=0$ to $t$, and noting that the body is initially at rest, we find the body velocity,
\begin{equation}
\dot{x}_1(t) = - \frac{m_2}{m_1+m_2} \dot{l}(t) + \frac{F_1 - F_2}{m_1 + m_2} t,
\end{equation}
and integrating again and setting $x_1(0) = 0$, we obtain the body position,
\begin{equation}
x_1(t) = - \frac{m_2}{m_1+m_2} l(t) + \frac{F_1 - F_2}{m_1 + m_2} \frac{t^2}{2}\cdot
\end{equation}
The first interval ends when the velocity of body 1 reaches  zero.  We expect this to occur very close to the end of the interval at a time $t = T - \Delta T$, where $\Delta T \ll T$ (see Fig.~\ref{onePhaseIntervals}).  We therefore expand $\dot{l}(t)$ near $t = T$ to find
\begin{equation}
\begin{split}
\dot{x}_1(T - \Delta T) = 0 = \Delta T \left [ m_2 \ddot{l}(T) - F_1 + F_2 \right ] + \left (F_1 - F_2 \right ) T + O(\Delta T^2), 
\end{split}
\end{equation}
which implies 
\begin{equation}
\Delta T = - \frac{F_1 - F_2}{m_2 \ddot{l}(T) - F_1 + F_2} T. \\
\end{equation}
To emphasize the dependence of quantities on the phase duration $T$, we note that if the length of extension remains constant we expect the kinematic acceleration to scale as $\ddot{l} \sim 1/T^2$.  We therefore introduce the notation $\ddot{l}(T) = l_T / T^2$, where $l_T$ is some constant which is proportional to the maximum separation between the bodies $l(T)$, and write
\begin{equation}
\Delta T = -\frac{F_1 - F_2}{m_2 l_T} T^3 + O(T^5).
\end{equation}
The position of body 1 at the end of the interval is
\begin{equation}
x_1(T-\Delta T) = - \frac{m_2}{m_1+m_2} l(T) + \frac{F_1 - F_2}{m_1 + m_2} \frac{T^2}{2} + O(T^3).
\end{equation}
We can now find the final body velocities and positions.  In the second interval we have that both body velocities are positive implying that $F_{f_1} = -F_1$ and $F_{f_2} = -F_2$.  This yields the equation of motion
\begin{equation}
\ddot{x}_1 = - \frac{m_2}{m_1+m_2} \ddot{l}(t) - \frac{F_1 + F_2}{m_1 + m_2}\cdot
\end{equation}
The velocity of body 1 is found by integration, where by specification the initial velocity of the body in this interval is zero,
\begin{equation}
\dot{x}_1(t) = - \frac{m_2}{m_1+m_2} \left [ \dot{l}(t) - \dot{l}(T-\Delta T) \right ] - \frac{F_1 + F_2}{m_1 + m_2} \left [ t - (T-\Delta T ) \right ],
\end{equation}
and at time $t = T$ we find
\begin{equation}
\dot{x}_1(T) = \frac{F_1 - F_2}{m_1 + m_2} T + O(T^3).
\end{equation}
It is not necessary to solve for the velocity of body 2, because by definition of the kinematics it will equal the velocity of body 1 at time $t = T$.  The position of body 1 is found by integrating the body velocity,
\begin{equation}
x_1(T) = - \frac{m_2}{m_1+m_2} l(T) + \frac{F_1 - F_2}{m_1 + m_2} \frac{T^2}{2} + O(T^4).
\end{equation}
Since $x_2(T) = x_1(T) + l(T)$, we find that
\begin{equation}
x_2(T) = \frac{m_1}{m_1+m_2} l(T) + \frac{1}{2} \frac{F_1 - F_2}{m_1 + m_2} T^2 + O(T^4), 
\end{equation}
and the geometric center $x_C = (x_1 + x_2)/2$ is
\begin{equation}\label{final}
x_C(T) = \frac{1}{2} \frac{m_1-m_2}{m_1 + m_2} l(T) + \frac{1}{2} \frac{F_1 - F_2}{m_1 + m_2} T^2 + O(T^4).
\end{equation}
The first term in this expression is the change in position due to the difference in body inertia and is independent of the phase duration.  The second term, however, is proportional to the difference in friction forces between the blocks and therefore contains the key to the direction of body translation.  Its origin is in the short interval of time that both bodies are moving forward: because an increase in phase duration causes the the length of this interval to increase, the change in final system position is correspondingly increased as well.

\subsubsection{A single extension phase when bodies have an initial positive velocity}

The situation detailed in the previous section is modified somewhat by the existence of an initial system velocity at the beginning of the kinematic period.  A similar asymptotic analysis (the details of which are given in Appendix A) with an initial positive velocity $\dot{x}_0$ yields the change in system velocity and position, respectively,
\begin{equation}
\dot{x}_C(T) - \dot{x}_0 = \frac{F_1 - F_2}{m_1 + m_2} T - \dot{x}_0 \frac{2 F_1}{m_2} \left ( \frac{1}{l_0} - \frac{1}{l_T} \right ) T^2 + O(T^3),
\end{equation}
and
\begin{equation}
x_C(T) = \frac{1}{2} \frac{m_1 - m_2}{m_1+m_2} l(T) + \dot{x}_0 T + \frac{1}{2} \frac{F_1 - F_2}{m_1 + m_2} T^2 - \dot{x}_0  \frac{2 F_1}{m_2 l_0} T^3 + O(T^4),
\end{equation}
where $l_T$ and $l_0$ are lengths which characterize the acceleration term in the kinematic specification.  The effect of the small initial velocity is to increase the total change in system position and to decrease the change in system velocity.  The first is intuitive and easy to understand.  The second is also intuitive and explains why the system exhibits stable dynamics, since period-wise increases in the initial velocity decay until the initial velocity reaches some steady-state.

\subsubsection{A full kinematic period when bodies translate in steady-state}

We can then address an extension phase followed by a contraction phase when the bodies are translating in steady-state. This requires us to first obtain the steady-state velocity for particular system parameters, and then insert this into our equation for the final body position after the full kinematic period.  To derive the steady-state velocity, we consider the change in velocity after an extension phase and a contraction phase and set the final and initial velocities equal to one another.  

For the contraction phase, the final velocity and position of the body when starting with some small initial positive velocity are simply the opposite of that for the extension phase,
\begin{equation}
\dot{x}_C(T) - \dot{x}_0 = -\frac{F_1 - F_2}{m_1 + m_2} T + \dot{x}_0 \frac{2 F_2}{m_1} \left ( \frac{1}{l_{0,c}} - \frac{1}{l_{T,c}} \right ) T^2 + O(T^3),
\label{contractionPhase}
\end{equation}
and
\begin{equation}
x_C(T) = -\frac{1}{2} \frac{m_1 - m_2}{m_1+m_2} l(T) + \dot{x}_0 T - \frac{1}{2} \frac{F_1 - F_2}{m_1 + m_2} T^2 + \dot{x}_0  \frac{2 F_2}{m_1 l_{0,c}} T^3 + O(T^4),
\end{equation}
where we have used the subscript $c$ to differentiate the kinematic constants in the contraction phase ($c$)  from the extension phase ($e$).  The details of this calculation are found in Appendix B.  For a contraction phase which is functionally equivalent and opposite to the extension phase, we have  $l_{0,e} = -l_{0,c}$, and $l_{T,e} = -l_{T,c}$, even if the phases have different durations.   We then find the steady-state velocity by combining an extension phase with a contraction phase and and setting the final velocity equal to the initial velocity.  By writing the extension phase duration as $\alpha T$ and the contraction phase duration as $(1-\alpha)T$ with kinematic period $T$, we obtain
\begin{equation}
\dot{x}_0 = \frac{(F_1 - F_2) \left ( 2 \alpha - 1 \right )}{2 T \left ( 1/l_0 - 1/l_T \right ) \left [ \alpha^2 F_1 m_1 + (1-\alpha)^2 F_2 m_2 \right ]} \left ( \frac{m_1 m_2}{m_1 + m_2} \right ) + O(T).
\end{equation}
We can simplify this expression by defining a reduced system mass $m^* = m_1 m_2 / (m_1 + m_2)$ and lumping the kinematic-dependent parameters into a single parameter with dimensions of length $L_{\text{kin}} = l_0 l_T / [2 (l_T - l_0)]$; if $L$ is the difference between the greatest and least separation of the  bodies, then $L_{kin} = L$ for a piecewise quadratic form for the kinematic specification and $\pi^2 L / 8$ for a sinusoidal form.  The relation becomes
\begin{equation}
\dot{x}_0 = (F_1 - F_2) \left ( 2 \alpha - 1 \right ) \frac{L_{\text{kin}} m^*}{\alpha^2 F_1 m_1 + (1-\alpha)^2 F_2 m_2} \frac{1}{T} + O(T).
\end{equation}
Accordingly we find that the steady-state initial velocity is inversely proportional to the total phase duration (or proportional to the prescribed kinematic velocity).  Using this initial velocity to compute the final position of the bodies,  we find that
\begin{equation}
x_C = (F_1 - F_2) \left ( 2 \alpha - 1 \right ) \frac{L_{\text{kin}} m^*}{\alpha^2 F_1 m_1 + (1-\alpha)^2 F_2 m_2} + O(T^4).
\label{finalx}
\end{equation}
This expression encapsulates the main result of this paper and explicitly establishes the two conditions necessary for locomotion.  The first condition is that $\alpha \ne 1/2$; or there must be an asymmetry between the extension and contraction phases of the kinematics.  The second condition is that $F_1 \ne F_2$; or there must be an asymmetry between the friction force exerted by each body on the supporting surface.  It is important to note that condition requires an asymmetry in the friction force specifically, regardless of how this force depends on material parameters (such as mass or friction coefficient).  Furthermore, we see that the change in position of the bodies therefore goes to a constant as the kinematic period vanishes, and the effective velocity of the system scales as $x_C / T \sim 1 / T$, or with the velocity of the kinematics.

\subsection{A physical description for all parameter regimes}

\begin{figure}[t]
\centering
\includegraphics[width = 1\textwidth]{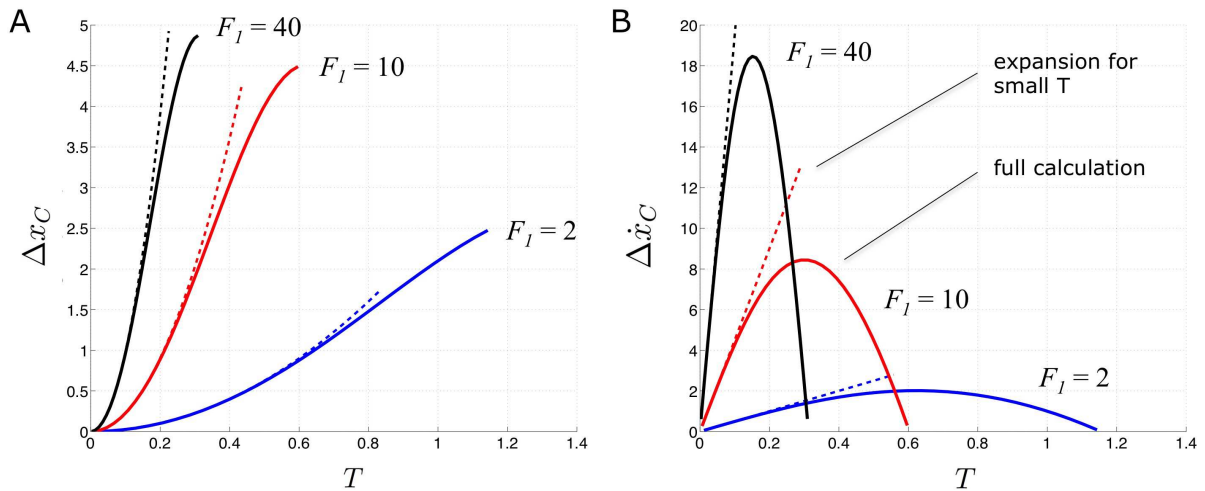}
\caption{Change in system position (A) and velocity (B) over a single extension phase versus phase duration;  (A): change in normalized system position, $\Delta x_C / L$;   (B) change in normalized system velocity, $\Delta\dot x_C \times T / L$.  The full calculation is plotted with solid lines while the asymptotic expansion presented in the text for small phase durations is shown with dashed lines. The parameters are $l(T) = 10$, $F_2 = 1$, $m_1 = m_2 = 1$ and the plots are shown for three different friction asymmetries $F_1 / F_2 = \{2, 10, 40\}$.}
\label{vel_pos_T}
\end{figure}

With an explanation in hand for when the driving force is large compared to the friction force, we now examine the relationship between phase interval and the change in system velocity and position for the full range of extension phase durations.  In Fig.~\ref{vel_pos_T} we plot the change in system  position (A) and velocity (B) against the length of the extension phase for all physically valid extension phase durations and a kinematic specification which is piecewise quadratic and continuous in position and velocity.  

The dependence on the change in system position on phase duration  (Fig.~\ref{vel_pos_T}A) provides an indication of the direction of system translation.  We see that this dependence is monotonic, which implies that the system will always translate forward when the extension phase is longer than the contraction phase, and vice versa.  

The dependence of system velocity on phase duration (Fig.~\ref{vel_pos_T}B)  provides an indication of the nature of the system dynamics.  Here we find that the change in system velocity is not monotonic with phase duration.  When the phase duration is short (large driving force, high Froude number) the change in system velocity increases as the phase duration increases.  When the phase duration is long (smaller driving force, relatively smaller Froude number) this dependence is reversed.

The combination of these two effects implies that there are two distinct regimes of translation for sliding locomotion of this system.  When the change in system velocity is increasing with increasing duration, period-wise initial velocity of the system will be positive when the bodies are translating in the forward direction.  However when the system velocity is decreases with increasing phase duration, we find that the period-wise initial velocity of the system is antagonistic to the overall direction of translation.  Finally, the maximum period-wise displacement of the system and therefore maximum average system velocity for a given kinematic specification occurs at the boundary between these two regimes. This can be explained by observing that a zero initial velocity at the beginning of the extension phase implies that the initial velocity for the contraction phase is at a maximum, which minimizes the decrease in system position over this phase.

\subsection{Parameter studies}

Once it is known that both an asymmetry in the specified kinematics and an asymmetry in the critical friction forces are necessary for the system to achieve net locomotion, we may ask how the motion of the system is dependent on the extent of the asymmetries.  One metric which characterizes the capacity of the system for locomotion is its velocity, and so to probe the effect of physical asymmetry on system behavior we define a ``period-averaged velocity'' as 
\begin{equation}
\langle U \rangle = \frac{x_C(t_0 + T) - x_C(t_0)}{T}\cdot
\end{equation}
For the kinematic specification, we choose a simple piecewise quadratic form and define a parameter $\alpha$ to characterize the asymmetry of the kinematics.  The length of the extension phase is then $\alpha T$ and the contraction phase is $(1-\alpha)T$, and $\alpha = 1/2$ implies a symmetric kinematic specification.

\begin{figure}[t]
\centering
\includegraphics[width = 1\textwidth]{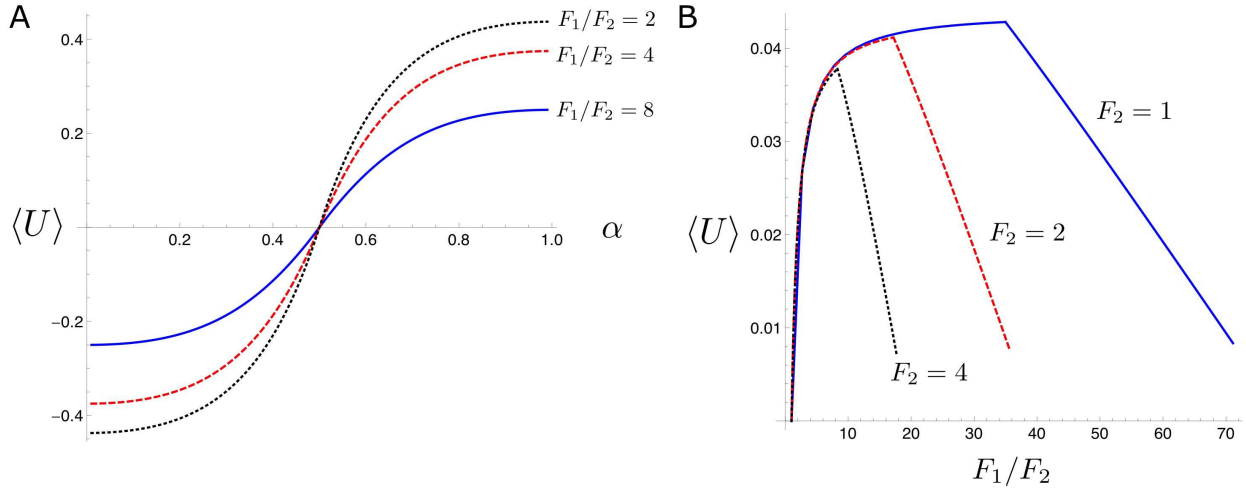}
\caption{(A):  Dependence of the period-averaged velocity, $\langle U \rangle$,  on the kinematic asymmetry,  $\alpha$, for $F_2 = 1$ and three values of $F_1$;  (B) Dependence of $\langle U \rangle$ on the friction asymmetry $F_1 / F_2$ for $\alpha = 3/4$ and three values of $F_2$. In both cases we have $m_1 = m_2 = 1$, $l(T) = 10$.}
\label{U_alpha_mu1}
\end{figure}

In Fig.~\ref{U_alpha_mu1}, we plot the period-averaged velocity against both the kinematic asymmetry (left) and the asymmetry in friction forces (right).  For the kinematic asymmetry, we observe that the period-averaged velocity of the system increases as the specified kinematics become more asymmetric.  

For the frictional asymmetry, we observe that the period-averaged velocity increases to a maximum and then decays to zero as the asymmetry increases, which is expected since no translation occurs either for $F_1/F_2 = 1$ or as $F_1 / F_2 \to \infty$ (when the friction force is too large for the kinematic specification to force the system into a sliding mode of locomotion).  Further, as explained in the previous section, we observe the two distinct regimes of translation in which the period-wise initial system velocity is either in the same direction as translation (small $F_1/F_2$) or in the opposite direction (large $F_1/F_2$).

\section{Discussion}

In this work, we attempt to understand the behavior of the conceptually simplest crawler possible: a one-dimensional, one-degree of freedom system consisting of two mechanically connected point masses.  We find that such a system is able to achieve time-averaged translation even when its frictional interactions with the surface are isotropic and are mediated only through sliding contact, in addition to the more evident cases of intermittent static contact and anisotropic friction.  We use computations and physical reasoning to show which and in what manner symmetries must to be broken to obtain net locomotion.  In doing so, we demonstrate that: 
\begin{itemize}
\item Friction-based locomotion with one degree of freedom is possible because of the non-linear dependence of the friction force on body velocity;
\item Two symmetries must be broken in order for the system to achieve locomotion: the two components of the body must exert a different friction force on the supporting surface, and the body kinematics must be asymmetric on a single period of actuation,
\item The physical mechanism of locomotion results from an interval within a single kinematic period in which the body with a smaller friction force (faster) carries the other body (slower) forward
\item For each chosen set of physical parameters and kinematic function the system always eventually achieves a steady-state of locomotion;
\item Two fundamental regimes of translation exist which correspond to the relative strength of the driving force to the friction force (or Froude number): when the driving force is large (higher Froude numbers) the system velocity at the beginning of each period is in the direction of translation; when the driving force is smaller (relatively lower Froude numbers) this initial system velocity opposes the direction of translation; and the maximum velocity occurs on the boundaries of these two regimes.
\end{itemize}

Further, we observe that locomotion in this system occurs necessarily at relatively high Froude number.  While experimental data on the forces exerted by crawling organisms is difficult to obtain, it seems likely that most crawling organisms locomote low Froude numbers.  This may have something to do with the energetic cost of maintaining high velocities (and therefore high Froude numbers) in frictional media; in our system, much mechanical energy is wasted in applying sufficiently large forces to the blocks to ensure that they always slide.  

The established criterion for the asymmetries of a two mass, friction-based system is fundamental knowledge in the physics of terrestrial locomotion.  Additionally we hope that this study will help guide the analysis and design of simple crawlers for use in exploration and medicine, as well aid in the development of intuition and understanding of more advanced and complex modes of terrestrial locomotion.

\section{Acknowledgments}

This work was supported by the National Science Foundation by Grant No.~CBET-0746285 to E.L. and a Focht-Powell Fellowship to G.L.W.

\appendix
\numberwithin{equation}{section}

\section{Analysis of a single extension phase when bodies start with some small initial velocity}

We examine here the effect some small initial positive velocity has on the final position and velocity of the system after a single extension phase.  When both bodies begin with a positive velocity the equation of motion for body 1 is
\begin{equation}
\ddot{x}_1 = - \frac{m_2}{m_1+m_2} \ddot{l} (t) - \frac{F_1 + F_2}{m_1 + m_2}\cdot
\end{equation}
Integration yields the body velocity,
\begin{equation}
\dot{x}_1(t) = - \frac{m_2}{m_1+m_2} \dot{l} (t) - \frac{F_1 + F_2}{m_1 + m_2} t + \dot{x}_0, 
\end{equation}
and the body position,
\begin{equation}
x_1(t) = - \frac{m_2}{m_1+m_2} l(t) - \frac{F_1 + F_2}{m_1 + m_2} \frac{t^2}{2} + \dot{x}_0 t,
\end{equation}
where we set the initial position of body 1 to $x_1(0) = 0$.  The first interval ends at time $t = t_1$ when the velocity of body 1 decreases to 0.  In general this can only be solved if $l(t)$ is specified, but to start let us assume that $t_1$ is small.  We can then expand $\dot{l}(t_1)$ to obtain
\begin{equation}
\dot{x}_1(t_1) = - \frac{m_2}{m_1+m_2} t_1 \ddot{l} (0) - \frac{F_1 + F_2}{m_1 + m_2} t_1 + \dot{x}_0 + O(t_1^2), 
\end{equation}
We then find that $\dot{x}_1(t_1) = 0$ when
\begin{equation}
t_1 = \dot{x}_0 \frac{m_1 + m_2}{m_2 \ddot{l} (0) + F_1 + F_2}\cdot
\end{equation}
When $T \to 0$ and using the notation $\ddot{l}(0) = l_0/T^2$ we have that
\begin{equation}
\begin{split}
t_1 &= \frac{\dot{x}_0 (m_1 + m_2)}{m_2 l_0}  T^2 \left ( \frac{1}{1 + T^2 \frac{F_1+F_2}{m_2 l_0}} \right ), \\
&= \frac{\dot{x}_0 (m_1 + m_2)}{m_2 l_0} T^2 \left [ 1 - T^2 \frac{F_1+F_2}{m_2 l_0} + T^4 \left ( \frac{F_1+F_2}{m_2 l_0} \right ) ^2 + ... \right ], \\
&= \dot{x}_0 \frac{(m_1 + m_2)}{m_2 l_0} T^2 \left ( 1 - T^2 \frac{F_1+F_2}{m_2 l_0} \right ) + O(T^6),
\end{split}
\end{equation}
which confirms our assumption that $t_1$ is small when $T \to 0$.  The position of body 1 at the end of the first interval is
\begin{equation}
\begin{split}
x_1(t_1) &= - \frac{m_2}{m_1+m_2} \frac{t_1^2}{2} \ddot{l} (0) - \frac{F_1 + F_2}{m_1 + m_2} \frac{t_1^2}{2} + \dot{x}_0 t_1, \\
&= t_1 \left ( \dot{x}_0 - \frac{t_1}{2} \frac{m_2 \ddot{l} (0) + F_1 + F_2}{m_1 + m_2} \right ), \\
&= \frac{\dot{x}_0 t_1}{2} = O(T^2).
\end{split}
\end{equation}
In the second interval, body 1 is moving backwards and body 2 is moving forwards.  The equation of motion is
\begin{equation}
\ddot{x}_1 = - \frac{m_2}{m_1+m_2} \ddot{l} (t) + \frac{F_1 - F_2}{m_1 + m_2}\cdot
\end{equation}
Integration yields the body velocity
\begin{equation}
\dot{x}_1(t) = - \frac{m_2}{m_1+m_2} \left [ \dot{l} (t) - \dot{l} (t_1) \right ] + \frac{F_1 - F_2}{m_1 + m_2} (t - t_1),
\end{equation}
and integrating again yields the body position
\begin{equation}
\begin{split}
x_1(t) &= - \frac{m_2}{m_1+m_2} \left [ l(t) - l(t_1) \right ] +  \frac{m_2}{m_1+m_2} \dot{l} (t_1) (t - t_1) \\
& \qquad + \frac{1}{2} \frac{F_1 - F_2}{m_1 + m_2} (t - t_1)^2 + x_1(t_1).
\end{split}
\end{equation}
This interval ends at time $t = t_2$ when the velocity of body 1 goes to zero again and both bodies are translating forwards.  We use the ansatz that $t_2$ is very close to the end of the interval $T$ and, as in the case where the bodies start from rest, define $\Delta T = T - t_2$.  We then have that
\begin{equation}
\begin{split}
\dot{l}(T - \Delta T) &= \dot{l} (T) - \Delta T \ddot{l}(T) + O(\Delta T^2), \\
&= - \Delta T \ddot{l}(T) + O(\Delta T^2), 
\end{split}
\end{equation}
and
\begin{equation}
\begin{split}
l(T - \Delta T) &= l(T) - \Delta T \dot{l}(T) + \frac{\Delta T^2}{2} \ddot{l}(T) + O(\Delta T^3), \\
&= \frac{\Delta T^2}{2} \ddot{l}(T) + O(\Delta T^3).
\end{split}
\end{equation}
The equation for body velocity becomes
\begin{equation}
\begin{split}
0 &= -\frac{m_2}{m_1 + m_2} \left [ - \Delta T \ddot{l}(T) - t_1 \ddot{l}(0) \right ] + \frac{F_1 - F_2}{m_1 + m_2} \left [ (T- \Delta T) - t_1 \right ], \\
&= \Delta T \left [ m_2 \ddot{l}(T) - (F_1 - F_2) \right ] + t_1 \left [ m_2 \ddot{l}(0) - (F_1 - F_2) \right ] \\
& \qquad + (F_1 - F_2) T, \\
&= \Delta T \left [ m_2 \ddot{l}(T) - (F_1 - F_2) \right ] + \dot{x}_0 (m_1 + m_2) \frac{m_2 \ddot{l}(0) - F_1 + F_2}{m_2 \ddot{l}(0) + F_1 + F_2} \\
& \qquad + (F_1 - F_2) T,
\end{split}
\end{equation}
This implies that
\begin{equation}
\begin{split}
\Delta T &= - \frac{1}{m_2 \ddot{l}(T) - F_1 + F_2} \Bigg [ \dot{x}_0 (m_1 + m_2) \frac{m_2 \ddot{l}(0) - F_1 + F_2}{m_2 \ddot{l}(0)  + F_1 + F_2} \cdots \\
& \qquad \cdots + (F_1 - F_2) T \Bigg ].
\end{split}
\end{equation}
Using the notation $\ddot{l}(0) = l_0/T^2$ and $\ddot{l}(T) = l_T/T^2$, we find
\begin{equation}
\begin{split}
\Delta T &= - \frac{T^2}{m_2 l_T} \left ( \frac{1}{1 - T^2\frac{F_1 - F_2}{m_2 l_T}} \right ) \Bigg [ \dot{x}_0 (m_1 + m_2) \frac{1 - T^2 \frac{F_1 - F_2}{m_2 l_0}}{1 + T^2 \frac{F_1 + F_2}{m_2 l_0}} \\
& \qquad + (F_1 - F_2) T \Bigg ], \\
&= - T^3 \frac{F_1-F_2}{m_2 l_T} \left ( 1 + T^2\frac{F_1 - F_2}{m_2 l_T} \right ) + O(T^7) ... \\
& \qquad - T^2 \frac{\dot{x}_0 (m_1 + m_2)}{m_2 l_T} \left ( 1 + T^2\frac{F_1 - F_2}{m_2 l_T} \right ) \left ( 1 - T^2\frac{F_1 - F_2}{m_2 l_0} \right ) \cdots \\
& \qquad \cdots \left ( 1 - T^2 \frac{F_1+F_2}{m_2 l_0} \right ) + O(T^6). \\
\end{split}
\end{equation}
Rearranging, we find
\begin{equation}
\begin{split}
\Delta T &= - \left [ \frac{\dot{x}_0 (m_1 + m_2)}{m_2 l_T} \right ] T^2 - \left (  \frac{F_1-F_2}{m_2 l_T} \right ) T^3 \\
& \qquad - \left [ \frac{\dot{x}_0 (m_1 + m_2)}{m_2 l_T} \left ( \frac{F_1 - F_2}{m_2 l_T} - \frac{2 F_1}{m_2 l_0} \right ) \right ] T^4 + O(T^5).
\end{split}
\end{equation}
\noindent The position of body 1 at the end of the second interval is
\begin{equation}
\begin{split}
x_1(T - \Delta T) &= -\frac{m_2}{m_1+m_2} \left [ l(T - \Delta T) - l(t_1) \right ] \\
& \qquad +  \frac{m_2}{m_1+m_2} \dot{l}(t_1) (T - \Delta T - t_1) \\
& \qquad + \frac{1}{2} \frac{F_1 - F_2}{m_1 + m_2} (T - \Delta T - t_1)^2 + x_1(t_1), \\
&= -\frac{m_2}{m_1+m_2} l(T) -\frac{1}{2} \frac{m_2}{m_1+m_2} \left [ \Delta T^2 \ddot{l}(T) - t_1^2 \ddot{l}(0) \right ] \\
& \qquad +  \frac{m_2}{m_1+m_2} t_1 \ddot{l}(0) (T - \Delta T - t_1) \\
& \qquad + \frac{1}{2} \frac{F_1 - F_2}{m_1 + m_2} (T - \Delta T - t_1)^2 + \frac{\dot{x}_0 t_1}{2} + O(T^4), \\
&=  -\frac{m_2}{m_1+m_2} l(T) + T^2 \left ( \frac{1}{2} \frac{F_1 - F_2}{m_1 + m_2} \right ) \\
& \qquad + \Delta T^2 \left [ - \frac{1}{2} \frac{m_2 \ddot{l}(T) }{m_1+m_2} +  \frac{1}{2} \frac{F_1 - F_2}{m_1 + m_2} \right ] \\
& \qquad + t_1^2 \left [ - \frac{1}{2} \frac{m_2 \ddot{l}(0) }{m_1+m_2} + \frac{1}{2} \frac{F_1 - F_2}{m_1 + m_2} \right ] \\ 
& \qquad - T \Delta T \left ( \frac{F_1 - F_2}{m_1 + m_2} \right ) + t_1 T \left [  \frac{m_2 \ddot{l}(0) }{m_1+m_2} - \frac{F_1 - F_2}{m_1 + m_2} \right ] \\
& \qquad + t_1 \Delta T \left [ - \frac{m_2 \ddot{l}(0) }{m_1+m_2} + \frac{F_1 - F_2}{m_1 + m_2} \right ] + \frac{\dot{x}_0 t_1}{2} + O(T^4).
\end{split}
\end{equation}
After much algebra, we obtain
\begin{equation}
\begin{split}
x_1(T-\Delta T) &= -\frac{m_2}{m_1+m_2} l(T) + \dot{x}_0 T \\
& \qquad + \frac{1}{2} \left [ \frac{F_1 - F_2}{m_1 + m_2} + (\dot{x}_0)^2 \frac{m_1 + m_2}{m_2 l_T} \right ] T^2 \\
& \qquad+ \frac{\dot{x}_0}{m_2} \left ( \frac{F_1 - F_2}{l_T} - \frac{2 F_1}{l_0} \right ) T^3 + O(T^4).
\end{split}
\end{equation}
In the final interval both bodies have a positive velocity.  The equation of motion for body 1 is 
\begin{equation}
\ddot{x}_1 = - \frac{m_2}{m_1+m_2} \ddot{l}(t) - \frac{F_1 + F_2}{m_1 + m_2}, \\
\end{equation}
so, noting that body 1 is at rest at the beginning of the final interval, the velocity of body 1 is
\begin{equation}
\dot{x}_1 = - \frac{m_2}{m_1+m_2} \left[ \dot{l}(t) - \dot{l}(t_2) \right ] - \frac{F_1 + F_2}{m_1 + m_2} (t - t_2),
\end{equation}
and its position,
\begin{equation}
\begin{split}
x_1 &= - \frac{m_2}{m_1+m_2} \left [ l(t) - l(t_2) \right ] + \frac{m_2}{m_1+m_2} \dot{l}(t_2) \left ( t - t_2 \right ) \\
& \qquad - \frac{1}{2} \frac{F_1 + F_2}{m_1 + m_2} (t - t_2)^2 + x_1(t_2).
\end{split}
\end{equation}
At time $t = T$ we find for the body velocity
\begin{equation}
\begin{split}
\dot{x}_1(T) &= - \frac{m_2}{m_1 + m_2} \left [ \dot{l}(T) - \dot{l}(T - \Delta T) \right ] - \frac{F_1 + F_2}{m_1 + m_2} \Delta T, \\
&= - \Delta T \left [ \frac{m_2}{m_1 + m_2} \ddot{l}(T) + \frac{F_1 + F_2}{m_1 + m_2} \right ] + O(T^4), \\
&= - \Delta T \left ( \frac{m_2}{m_1 + m_2} l_T T^{-2} + \frac{F_1 + F_2}{m_1 + m_2} \right ) + O(T^4), \\
&= \dot{x}_0 + \frac{F_1 - F_2}{m_1 + m_2} T - \dot{x}_0 \frac{2 F_1}{m_2} \left ( \frac{1}{l_0} - \frac{1}{l_T} \right ) T^2 + O(T^3),
\end{split}
\end{equation}
and for the position of body 1,
\begin{equation}
\begin{split}
x_1 &= - \frac{m_2}{m_1+m_2} \left [ l(T) - l(T - \Delta T) \right ] + \frac{m_2}{m_1+m_2} \dot{l}(T - \Delta T) \Delta T \\
& \qquad - \frac{1}{2} \frac{F_1 + F_2}{m_1 + m_2} \Delta T^2 + x_1(T - \Delta T), \\
&=  - \frac{1}{2} \frac{m_2}{m_1+m_2} \Delta T^2 \ddot{l}(T) - \frac{1}{2} \frac{F_1 + F_2}{m_1 + m_2} \Delta T^2 + x_1(T - \Delta T) + O(T^6), \\
&= - \frac{1}{2} \frac{m_2} {m_1+m_2} \frac{l_T \Delta T^2}{T^2} + x_1(T-\Delta T) + O(T^4), \\
&= -\frac{m_2}{m_1+m_2} l(T) + \dot{x}_0 T + \frac{1}{2} \left ( \frac{F_1 - F_2}{m_1 + m_2} \right ) T^2 + \dot{x}_0 \frac{2 F_1}{m_2 l_0} T^3 + O(T^4).
\end{split}
\end{equation}
At the end of the extension phase, the velocity of body 2 is equal to that of body 1, and the position of body 2 is $x_2(T) = x_1(T) + l(T)$.  The velocity of the geometric center $\dot{x}_C = (\dot{x}_1 + \dot{x}_2)/2$ is therefore
\begin{equation}
\dot{x}_C(T) = \dot{x}_0 + \frac{F_1 - F_2}{m_1 + m_2} T - \dot{x}_0 \frac{2 F_1}{m_2} \left ( \frac{1}{l_0} - \frac{1}{l_T} \right ) T^2 + O(T^3),
\end{equation}
and the position of the geometric center $x_C = (x_1+x_2)/2$ is
\begin{equation}
x_C(T) = \frac{1}{2} \frac{m_1 - m_2}{m_1+m_2} l(T) + \dot{x}_0 T + \frac{1}{2} \left ( \frac{F_1 - F_2}{m_1 + m_2} \right ) T^2 + \dot{x}_0 \frac{2 F_1}{m_2 l_0} T^3 + O(T^4).
\end{equation}

\section{Analysis of a single contraction phase when bodies start with some small initial velocity}

We examine here the change in velocity of the two body system after a contraction phase when the bodies begin and end with some small initial positive velocity.  This velocity is presumed small enough such that the contraction kinematics induce a reversal in the velocity of body 2, but large enough such that the body still has a positive velocity at the end of the phase.  This analysis is very similar to the extension phase analysis, except that it is the velocity of body 2 which is now of interest.  The equation of motion for body 2 is
\begin{equation}
\ddot{x}_2 = \frac{m_1}{m_1+m_2} \ddot{l} (t) - \frac{F_1 + F_2}{m_1 + m_2}\cdot
\end{equation}
This equation of motion holds until the velocity of body 2 changes sign.  Using the same steps as for the extension phase and letting $T \to 0$, we find the time $t_1$ at which $\dot{x}_2(t_1) = 0$,
\begin{equation}
\begin{split}
t_1 &= \dot{x}_0 \frac{m_1 + m_2}{-m_1 \ddot{l}(0) + F_1 + F_2}, \\
&= - \dot{x}_0 \frac{m_1+m_2}{m_1 l_0} T^2 \left ( 1 + T^2 \frac{F_1 + F_2}{m_1 l_0} \right ) + O(T^6).
\end{split}
\end{equation}
The position of body 2 at $t = t_1$ is
\begin{equation}
\begin{split}
x_2(t_1) &= x_2(0) + \frac{m_1}{m_1+m_2} \left ( l(t_1) - l(0) \right ) - \frac{1}{2} \frac{F_1 - F_2}{m_1 + m_2} t_1^2 + \dot{x}_0 t_1, \\
&= x_2(0) + t_1 \left [ \dot{x}_0 - \frac{t_1}{2} \frac{F_1 + F_2 - \ddot{l}(0) m_1}{m_1 + m_2} \right ], \\
&= x_2(0) + \frac{\dot{x}_0 t_1}{2} + O(T^2).
\end{split}
\end{equation}
In the second interval, body 1 is moving forwards and body 2 is moving backwards, and the equation of motion for body 2 is
\begin{equation}
\ddot{x}_2 = \frac{m_1}{m_1+m_2} \ddot{l}(t) - \frac{F_1-F_2}{m_1+m_2}\cdot
\end{equation}
As in the extension phase with body 1, this interval ends when the velocity of body 2 changes sign.  It is convenient to again define $\Delta T = T - t_2$, and we find 
\begin{equation}
\begin{split}
\Delta T &= \frac{1}{m_1 \ddot{l}(T) - F_1 + F_2} \Bigg [ \dot{x}_0 \left ( m_1 + m_2 \right ) \frac{m_1 \ddot{l}(0) - F_1 + F_2}{m_1 \ddot{l}(0) - F_1 - F_2 } \\
& \qquad - (F_1 - F_2) T \Bigg ].
\end{split}
\end{equation}
With the notation $\ddot{l}(0) = l_0/T^2$ and $\ddot{l}(T) = l_T/T^2$, and expanding in powers of $T$, this becomes
\begin{equation}
\begin{split}
\Delta T &= -\frac{F_1 - F_2}{m_1 l_T} T^3 \left ( 1 + T^2 \frac{F_1 - F_2}{m_1 l_T} \right ) + O(T^7) \\
& \qquad + T^2 \frac{\dot{x}_0 (m_1 + m_2 )}{m_1 l_T} \left ( 1 + T^2 \frac{F_1 - F_2}{m_1 l_T} \right ) \left ( 1 - T^2 \frac{F_1 - F_2}{m_1 l_0} \right ) \cdots \\
& \qquad \cdots \left ( 1 + T^2 \frac{F_1 + F_2}{m_1 l_0} \right ) + O(T^6), \\
&= \left [ \frac{\dot{x}_0 (m_1 + m_2 )}{m_1 l_T} \right ] T^2 - \left ( \frac{F_1 - F_2}{m_1 l_T} \right ) T^3 \\
& \qquad + \left [ \frac{\dot{x}_0 (m_1 + m_2) }{m_1 l_T} \left ( \frac{F_1 - F_2}{m_1 l_T} + \frac{2 F_2}{m_1 l_0} \right ) \right ] T^4 + O(T^5).
\end{split}
\end{equation}
Expanding the kinematic specification in the same way as in the extension phase calculation, and noting that in this case $l(T) = 0$, we find the position of body 2 at the end of the second interval,
\begin{equation}
\begin{split}
x_2(T - \Delta T) &=  \frac{m_1}{m_1+m_2} \left [ l(T-\Delta T) - l(t_1) \right ] \\
& \qquad - \frac{m_1}{m_1+m_2} \dot{l}(t_1) \left ( T - \Delta T - t_1 \right ) \\
& \qquad - \frac{1}{2} \frac{F_1-F_2}{m_1+m_2} \left ( T - \Delta T - t_1 \right )^2 + x_2(t_1), \\
&= - \frac{m_1}{m_1+m_2} l(0) + \frac{1}{2} \frac{m_1}{m_1+m_2} \left [ \Delta T^2 \ddot{l}(T) - t_1^2 \ddot{l}(0) \right ] \\
& \qquad -\frac{m_1}{m_1+m_2} t_1 \ddot{l}(0) \left ( T - \Delta T - t_1 \right ) \\
& \qquad - \frac{1}{2} \frac{F_1-F_2}{m_1+m_2} \left ( T - \Delta T - t_1 \right )^2 + x_2(0) + \frac{\dot{x}_0 t_1}{2} + O(T^4), \\
&= - \frac{m_1}{m_1+m_2} l(0) - T^2 \left ( \frac{1}{2} \frac{F_1-F_2}{m_1+m_2} \right ) \\
& \qquad - t_1 T \left [ \frac{m_1 \ddot{l}(0)}{m_1+m_2} - \frac{F_1-F_2}{m_1+m_2} \right ] + T \Delta T \left ( \frac{F_1-F_2}{m_1+m_2} \right ) \\
& \qquad + \Delta T^2 \left [ \frac{1}{2} \frac{m_1 \ddot{l}(T)}{m_1+m_2} - \frac{1}{2} \frac{F_1-F_2}{m_1+m_2} \right ] \\
& \qquad + t_1^2 \left ( \frac{m_1 \ddot{l}(0)}{m_1+m_2} - \frac{1}{2} \frac{F_1-F_2}{m_1+m_2} \right ) \\
& \qquad + t_1 \Delta T \left [ \frac{m_1 \ddot{l}(0)}{m_1+m_2} - \frac{F_1-F_2}{m_1+m_2} \right ] \\
& \qquad + x_2(0) + \frac{\dot{x}_0 t_1}{2} + O(T^4), \\
\end{split}
\end{equation}
which simplifies to
\begin{equation}
\begin{split}
x_2(T-\Delta T) &= x_2(0) - \frac{m_1}{m_1+m_2} l(0) + \dot{x}_0 T \\
& \qquad - \frac{1}{2} \left [ \frac{F_1-F_2}{m_1+m_2} + (\dot{x}_0)^2  \frac{m_1+m_2}{m_1 l_T} \right ] T^2 \\
& \qquad + \frac{\dot{x}_0}{m_1} \left ( \frac{F_1 - F_2}{l_T} + \frac{2 F_2}{l_0} \right ) T^3 + O(T^4) \cdot
\end{split}
\end{equation} 
In the third interval, both bodies are translating forward.  The equation of motion for body 2 is then
\begin{equation}
\ddot{x}_2 = \frac{m_1}{m_1+m_2} \ddot{l} (t) - \frac{F_1 + F_2}{m_1 + m_2}\cdot
\end{equation}
By integrating from $T - \Delta T$ to $T$ and employing the fact that $\dot{x}_2(T-\Delta T) = 0$, we find the velocity of body 2 at the end of the contraction phase,
\begin{equation}
\begin{split}
\dot{x}_2(T) &= \frac{m_1}{m_1+m_2} \left [ \dot{l}(T) - \dot{l}(T-\Delta T) \right ] - \frac{F_1 + F_2}{m_1 + m_2} \Delta T, \\
&=  \frac{m_1}{m_1+m_2} \Delta T \ddot{l}(T) - \frac{F_1 + F_2}{m_1 + m_2} \Delta T, \\
&= \dot{x}_0 - \left ( \frac{F_1 - F_2}{m_1+m_2} \right ) T + \dot{x}_0 \frac{2 F_2}{m_1} \left ( \frac{1}{l_0} - \frac{1}{l_T} \right ) + O(T^3) .
\end{split}
\end{equation}
The position of body 2 is
\begin{equation}
\begin{split}
x_2(T) &=  \frac{m_1}{m_1+m_2} \left [ l(T) - l(T-\Delta T) \right ] + \frac{m_1}{m_1+m_2} \dot{l}(T - \Delta T) \\
& \qquad - \frac{1}{2}  \frac{F_1 + F_2}{m_1 + m_2} \Delta T ^2 + \dot{x}_2(T-\Delta T), \\
&= \frac{1}{2} \frac{m_1}{m_1+m_2} \Delta T^2 \ddot{l}(T) + \dot{x}_2(T-\Delta T) + O(T^4), \\
&= x_2(0) - \frac{m_1}{m_1+m_2} l(0) + \dot{x}_0 T - \frac{1}{2} \frac{F_1-F_2}{m_1+m_2} T^2 \\
& \qquad + \dot{x}_0 \frac{2 F_2}{m_1 l_0} T^3 + O(T^4).
\end{split}
\end{equation}
Finally, we find the velocity of the geometric center by observing that at the end of the contraction phase we have $\dot{x}_1(T) = \dot{x}_2(T) = \dot{x}_C(T)$, and we find the position of the geometric center by observing that because $l(T) = 0$ then we must have $x_C(T) = x_2(T)$.  Further, if we presume that, as in the extension interval calculation, that the initial position of the geometric center is $x_C(0) = 0$, we find that $x_2(0) = l(0)/2$.  The velocity of the geometric center at the end of the contraction phase is therefore
\begin{equation}
\dot{x}_C(T) = \dot{x}_0 - \left ( \frac{F_1 - F_2}{m_1+m_2} \right ) T + \dot{x}_0 \frac{2 F_2}{m_1} \left ( \frac{1}{l_0} - \frac{1}{l_T} \right ) + O(T^3),
\end{equation}
and the position of the geometric center is
\begin{equation}
x_C(T) = - \frac{1}{2} \frac{m_1 - m_2}{m_1+m_2} l(0) + \dot{x}_0 T - \frac{1}{2} \frac{F_1-F_2}{m_1+m_2} T^2 + \dot{x}_0 \frac{2 F_2}{m_1 l_0} T^3 + O(T^4).
\end{equation}

\pagebreak
\bibliographystyle{model2-names}
\bibliography{refsArticles,refsBooks}

\begin{thebibliography}{22}
\expandafter\ifx\csname natexlab\endcsname\relax\def\natexlab#1{#1}\fi
\expandafter\ifx\csname url\endcsname\relax
  \def\url#1{\texttt{#1}}\fi
\expandafter\ifx\csname urlprefix\endcsname\relax\def\urlprefix{URL }\fi
\providecommand{\eprint}[2][]{\url{#2}}
\providecommand{\bibinfo}[2]{#2}
\ifx\xfnm\relax \def\xfnm[#1]{\unskip,\space#1}\fi
%Type = Book
\bibitem[{Alexander(2003)}]{alexander2003}
\bibinfo{author}{Alexander, R.}, \bibinfo{year}{2003}.
\newblock \bibinfo{title}{Principles of animal locomotion}.
\newblock \bibinfo{publisher}{Princeton University Press}.
%Type = Book
\bibitem[{Bender and Orszag(1999)}]{benderOrszag1999}
\bibinfo{author}{Bender, C.}, \bibinfo{author}{Orszag, S.},
  \bibinfo{year}{1999}.
\newblock \bibinfo{title}{Advanced Mathematical Methods for Scientists and
  Engineers I: Asymptotic Methods and Perturbation Theory}.
\newblock \bibinfo{publisher}{Springer-Verlag}.
%Type = Article
\bibitem[{Berrigan and Pepin(1995)}]{berriganPepin1995}
\bibinfo{author}{Berrigan, D.}, \bibinfo{author}{Pepin, D.J.},
  \bibinfo{year}{1995}.
\newblock \bibinfo{title}{How maggots move: Allometry and kinematics of
  crawling in larval diptera}.
\newblock \bibinfo{journal}{J. Insect Physiol.} \bibinfo{volume}{41},
  \bibinfo{pages}{329--337}.
%Type = Article
\bibitem[{Bolotnik and et~al.(2011)}]{bolotnik2011}
\bibinfo{author}{Bolotnik, N.}, \bibinfo{author}{et~al.}, \bibinfo{year}{2011}.
\newblock \bibinfo{title}{The undulatory motion of a chain of particles in a
  resistive medium}.
\newblock \bibinfo{journal}{Z. Angew. Math. Mech.} \bibinfo{volume}{91}.
%Type = Article
\bibitem[{Chernous'ko(2002)}]{chernousko2002}
\bibinfo{author}{Chernous'ko, F.L.}, \bibinfo{year}{2002}.
\newblock \bibinfo{title}{The optimum rectilinear motion of a two-mass system}.
\newblock \bibinfo{journal}{J. Appl. Math. Mech-USS} \bibinfo{volume}{66}.
%Type = Article
\bibitem[{Figurina(2004)}]{figurina2004}
\bibinfo{author}{Figurina, T.Y.}, \bibinfo{year}{2004}.
\newblock \bibinfo{title}{Controlled quasistatic motions of a two-link robot on
  a horizontal plane}.
\newblock \bibinfo{journal}{J. Comput. Sys. Sci.} \bibinfo{volume}{43}.
%Type = Article
\bibitem[{Gans(1975)}]{gans1975}
\bibinfo{author}{Gans, C.}, \bibinfo{year}{1975}.
\newblock \bibinfo{title}{Tetrapod limblessness: Evolution and functional
  corollaries}.
\newblock \bibinfo{journal}{Am. Zool.} \bibinfo{volume}{15},
  \bibinfo{pages}{455--467}.
%Type = Article
\bibitem[{Gray(1946)}]{gray1946}
\bibinfo{author}{Gray, J.}, \bibinfo{year}{1946}.
\newblock \bibinfo{title}{The mechanism of locomotion in snakes}.
\newblock \bibinfo{journal}{J. Exp. Biol.} \bibinfo{volume}{23}.
%Type = Article
\bibitem[{Hirose and Morishima(1990)}]{hirose1990}
\bibinfo{author}{Hirose, S.}, \bibinfo{author}{Morishima, A.},
  \bibinfo{year}{1990}.
\newblock \bibinfo{title}{Design and control of a mobile robot with an
  articulated body}.
\newblock \bibinfo{journal}{Int. J. Robot. Res.} \bibinfo{volume}{9}.
%Type = Article
\bibitem[{Home(1812)}]{home1812}
\bibinfo{author}{Home, E.}, \bibinfo{year}{1812}.
\newblock \bibinfo{title}{Observations intended to show that the progressive
  motion of snakes is partly performed by means of the ribs}.
\newblock \bibinfo{journal}{Philos. Trans} \bibinfo{volume}{163}.
%Type = Article
\bibitem[{Hu and et~al.(2009)}]{hu2009}
\bibinfo{author}{Hu, D.L.}, \bibinfo{author}{et~al.}, \bibinfo{year}{2009}.
\newblock \bibinfo{title}{The mechanics of slithering locomotion}.
\newblock \bibinfo{journal}{P. Natl. Acad. Sci. USA} \bibinfo{volume}{106}.
%Type = Article
\bibitem[{Jing and Alben(2012)}]{jingAlben2012}
\bibinfo{author}{Jing, F.}, \bibinfo{author}{Alben, S.}, \bibinfo{year}{2012}.
\newblock \bibinfo{title}{Optimization of two- and three-link snake-like
  locomotion}.
\newblock \bibinfo{journal}{ArXiv} \bibinfo{volume}{1212.0062v1}.
%Type = Article
\bibitem[{Lauga and Powers(2009)}]{lauga09}
\bibinfo{author}{Lauga, E.}, \bibinfo{author}{Powers, T.R.},
  \bibinfo{year}{2009}.
\newblock \bibinfo{title}{The hydrodynamics of swimming microorganisms}.
\newblock \bibinfo{journal}{Rep. Prog. Phys.} \bibinfo{volume}{72},
  \bibinfo{pages}{096601}.
%Type = Article
\bibitem[{Mosauer(1932)}]{mosauer1932}
\bibinfo{author}{Mosauer, W.}, \bibinfo{year}{1932}.
\newblock \bibinfo{title}{On the locomotion of snakes}.
\newblock \bibinfo{journal}{Science} \bibinfo{volume}{76}.
%Type = Article
\bibitem[{Purcell(1977)}]{purcell1977}
\bibinfo{author}{Purcell, E.M.}, \bibinfo{year}{1977}.
\newblock \bibinfo{title}{Life at low {R}eynolds number}.
\newblock \bibinfo{journal}{Am. J. Phys.} \bibinfo{volume}{45}.
%Type = Article
\bibitem[{Quillin(1999)}]{quillin1999}
\bibinfo{author}{Quillin, K.J.}, \bibinfo{year}{1999}.
\newblock \bibinfo{title}{Kinematic scaling of locomotion by hydrostatic
  animals: ontogeny of peristaltic crawling by the earthworm lumbricus
  terrestris}.
\newblock \bibinfo{journal}{J. Exp. Biol.} \bibinfo{volume}{202},
  \bibinfo{pages}{661--674}.
%Type = Article
\bibitem[{Transeth et~al.(2009)Transeth, Pettersen and
  Liljeback}]{transeth2009}
\bibinfo{author}{Transeth, A.A.}, \bibinfo{author}{Pettersen, K.Y.},
  \bibinfo{author}{Liljeback, P.}, \bibinfo{year}{2009}.
\newblock \bibinfo{title}{A survey on snake robot modeling and locomotion}.
\newblock \bibinfo{journal}{Robotica} \bibinfo{volume}{27},
  \bibinfo{pages}{999--1015}.
%Type = Book
\bibitem[{Vogel(1994)}]{vogel1994}
\bibinfo{author}{Vogel, S.}, \bibinfo{year}{1994}.
\newblock \bibinfo{title}{Life in moving fluids}.
\newblock \bibinfo{publisher}{Princeton University Press}.
%Type = Article
\bibitem[{Walton and et~al.(1991)}]{walton1991}
\bibinfo{author}{Walton, M.}, \bibinfo{author}{et~al.}, \bibinfo{year}{1991}.
\newblock \bibinfo{title}{The energetic cost of limbless locomotion}.
\newblock \bibinfo{journal}{Science} \bibinfo{volume}{249}.
%Type = Article
\bibitem[{Zimmerman and et~al.(2004)}]{zimmerman2004}
\bibinfo{author}{Zimmerman, K.}, \bibinfo{author}{et~al.},
  \bibinfo{year}{2004}.
\newblock \bibinfo{title}{An approach to worm-like motion}.
\newblock \bibinfo{journal}{XXI ICTAM, 15-21 August, Warsaw, Poland} .
%Type = Article
\bibitem[{Zimmerman and et~al.(2007)}]{zimmerman2007b}
\bibinfo{author}{Zimmerman, K.}, \bibinfo{author}{et~al.},
  \bibinfo{year}{2007}.
\newblock \bibinfo{title}{Forced nonlinear oscillator with nonsymmetric dry
  friction}.
\newblock \bibinfo{journal}{Arch. Appl. Mech.} \bibinfo{volume}{77}.
%Type = Article
\bibitem[{Zimmerman and et~al.(2009)}]{zimmerman2009}
\bibinfo{author}{Zimmerman, K.}, \bibinfo{author}{et~al.},
  \bibinfo{year}{2009}.
\newblock \bibinfo{title}{Motion of a chain of three point masses on a rough
  plane under kinematical constraints}.
\newblock \bibinfo{journal}{Modeling, Simulation and Control of Nonlinear
  Engineering Dynamical Systems} , \bibinfo{pages}{61--70}.

\end{thebibliography}

\end{document}